\documentstyle[proceedings,natbibkap,psfig]{crckapb} 

\def\ifundefined#1{\expandafter\ifx\csname#1\endcsname\relax}

\def\la{\mathrel{\hbox{\rlap{\hbox{\lower4pt\hbox{$\sim$}}}\hbox{$<$}}}}
\def\ga{\mathrel{\hbox{\rlap{\hbox{\lower4pt\hbox{$\sim$}}}\hbox{$>$}}}}

\newcommand{\be}{\begin{eqnarray}}
\newcommand{\ee}{\end{eqnarray}}

\ifundefined{nuc}\def\nuc#1#2{\relax\ifmmode{}^{#1}{\protect\text{#2}}%
\else${}^{#1}$#2\fi}\else\relax\fi

\newcommand{\etal}{et al.}

\newcommand{\kmps}{km~s$^{-1}$}

\newcommand{\msol}{\ifmmode{{\rm M}_\odot}\else{M$_\odot$}\fi}
\newcommand{\foe}{\ifmmode{10^{51}}\else{$10^{51}$}\fi}

\newcommand{\xni}{\ifmmode{{\rm X}_{\rm Ni}}\else{X$_{\rm Ni}$}\fi}

\def\Teff{\ifmmode{T_{\rm eff}}\else{\hbox{$T_{\rm eff}$} }\fi}
\def\Rzero{\ifmmode{R_0}\else{\hbox{$R_0$} }\fi}
\newcommand{\vno}{\ifmmode{v_0}\else{\hbox{$v_0$} }\fi}

\newcommand{\inu}{\ifmmode{I_\nu}\else{$I_\nu$}\fi}
\newcommand{\jnu}{\ifmmode{J_\nu}\else{$J_\nu$}\fi}
\newcommand{\hnu}{\ifmmode{H_\nu}\else{$H_\nu$}\fi}
\newcommand{\knu}{\ifmmode{K_\nu}\else{$K_\nu$}\fi}

\newcommand{\mnras}{MNRAS}

\begin{opening}
\title{NLTE Modeling of SNe Ia Near Maximum Light}
\author{E.~Baron}
\institute{Dept. of Physics and Astronomy, University of Oklahoma, 440 W.
Brooks, Rm 131, Norman, OK 73019-0225, USA.}
\author{P.~H.~Hauschildt}
\institute{Dept. of Physics and Astronomy, Arizona State
University, Tempe, AZ 85287-1504, USA.}
\author{A.~Mezzacappa}
\institute{Theoretical and Computational Physics Group, Physics
Division, Oak Ridge National Laboratory, Oak Ridge, TN 37831-6373, USA\\ and\\
Dept. of Physics and Astronomy, 401 A. H. Nielsen
Physics Building, University of Tennessee, 
Knoxville, TN 37996-1200, USA.}
\end{opening}

\begin{document}

\bibliographystyle{natbib-apj}

\citestyle{kluwer}

\begin{abstract}
Modeling the atmospheres of SNe~Ia requires the solution of the
NLTE radiative transfer equation. We discuss the formulation of the
radiative transfer equation in the co-moving frame. For characteristic
velocities larger than $\sim 2000$~\kmps, the effects of advection on
the synthetic spectra are non-negligible, and hence should be
included in model calculations. We show that the time-independent or
quasi-static approximation is adequate for SNe~Ia near maximum light,
as well as for most other astrophysical problems; e.g., hot stars,
novae, and other types of supernovae. We examine the use of the
Sobolev approximation in modeling moving atmospheres and find that the
number of overlapping lines in the co-moving frame make the
approximation suspect in models that predict both lines and continua.
We briefly discuss the form of the Rosseland mean opacity in the
co-moving frame, and present a formula that is easy to implement in
radiation hydrodynamics calculations.
\end{abstract}

\section{Introduction}

There are many astrophysical systems that require a solution to
the radiative transfer equation in moving media; e.g., Wolf-Rayet
and other hot stars with stellar winds, novae, supernovae, the
material surrounding 
quasars, and even the early phases of the universe when the
material is still optically thick \cite[]{mih80}. Because there is such
a large simplification in the radiation-matter interaction terms, it
is both customary and expedient to solve the radiation transfer
equation in the co-moving frame.  In a series of papers Mihalas and
co-workers \cite[]{mkh75,mkh76a,mkh76b,mkh76c,mk78} examined methods
for solving the co-moving frame line-transfer problem, where the
Doppler effect dominates because the characteristic width
over which the line profile varies is small.  This effectively
increases the importance of the Doppler effect over the other $O(v/c)$
(advection and aberration) effects by the ratio $c/v_{\rm therm}$, where
$v_{\rm therm}$ is the thermal velocity corresponding to the intrinsic
Doppler line-width. Recently, there have been increasingly
sophisticated attempts to model the atmospheres of hot stars
\cite[e.g.,][]{werner87},
novae \cite[]{hsawss94,phhnov95}, and supernovae
\cite[]{bran91,eastpin93,hofsn94d,nug1a95,b93j3}, including  NLTE for
lines and continua,
and the effects
of line blanketing. Here we systematically discuss the
important effects that must be included when solving the radiative
transfer equation in the co-moving frame, elucidate the range of
applicability of Eulerian approximations such as the Sobolev
approximation, and  present a co-moving formulation of the Rosseland
mean opacity. Some of the results discussed here are also discussed in
\cite{bhm95}.

\section{Radiative Transfer Equation}

The co-moving frame radiative transfer equation for spherically
symmetric flows can be written 
 as \cite[cf.][]{found84}:

\vbox{\be
&\quad&\gamma (1+\beta\mu)\frac{\partial\inu}{\partial t} + \gamma (\mu +
\beta) \frac{\partial\inu}{\partial r}\nonumber\\
& +& \frac{\partial}{\partial
\mu}\left\{ \gamma (1-\mu^2)\left[ \frac{1+\beta\mu}{r}
\right.\right.\nonumber\\
&\quad&\left.\left. \quad -\gamma^2(\mu+\beta)
\frac{\partial\beta}{\partial r} -  
\gamma^2(1+\beta\mu) \frac{\partial\beta}{\partial
t}\right] \inu\right\} \nonumber\\
&-&  \frac{\partial}{\partial
\nu}\left\{ \gamma\nu\left[ \frac{\beta(1-\mu^2)}{r}
+\gamma^2\mu(\mu+\beta) \frac{\partial\beta}{\partial r}
\right.\right.\nonumber\\
&\quad& \left.\left.\quad  +
\gamma^2\mu(1+\beta\mu) \frac{\partial\beta}{\partial
t}\right]\inu\right\}\label{fullrte}\\
&+&\gamma\left\{\frac{2\mu+\beta(3-\mu^2)}{r}\right.
\nonumber\\
&\quad&\quad
\left. +\gamma^2(1+\mu^2+2\beta\mu)\frac{\partial\beta}{\partial r} + 
\gamma^2[2\mu + \beta(1+\mu^2)]\frac{\partial\beta}{\partial
t}\right\}\inu \nonumber\\
&\quad& = \eta_\nu - \chi_\nu\inu.\nonumber
\ee}
\noindent
We set  $c=1$; $\beta$ is the velocity; and
$\gamma = (1-\beta^2)^{-1/2}$ is the usual Lorentz factor. We emphasize
that, in Eq.~\ref{fullrte},  the physical (dependent) variables are
all evaluated in the co-moving Lagrangian frame. However, the choice of
independent variables is free, and the coordinate $r$ in
Eq.~\ref{fullrte} is an {\em Eulerian\/} variable \cite[for a discussion
of this point, cf.][]{mezzmat89}. This is the most
convenient choice for solving the transfer equation, where one usually
specifies the grid by fixing the optical depth for some reference
frequency. However, this grid differs from the fully Lagrangian grid
typically used in radiation hydrodynamics. In the latter case,
$r\equiv r(m)$.

In order to illuminate the physics, and without loss of generality, we
expand Eq.~\ref{fullrte} in powers of $\beta$ and keep terms only to
$O(\beta)$. While this is not necessary \cite[]{mkh76b,mih80,phhs392}, it is
adequate for most astrophysical flows. To $O(\beta)$, the radiation
transport equation becomes: 
\be
 \frac{\partial\inu}{\partial t} +
(\mu + \beta) \frac{\partial\inu}{\partial r}\nonumber\\ 
+(1-\mu^2)\left[ \mu
(\frac{\beta}{r} - \frac{\partial\beta}{\partial r}) +
\frac{1}{r}-\frac{\partial\beta}{\partial t}
\right]\frac{\partial\inu}{\partial \mu}\nonumber\\
+ \left[
\mu^2(\frac{\beta}{r} - \frac{\partial\beta}{\partial r}) -
\frac{\beta}{r}\right]\left\{\frac{\partial\inu}{\partial \ln\nu} -
3\inu\right\} - \mu\frac{\partial\beta}{\partial
t}\frac{\partial\inu}{\partial \ln\nu}\label{rteob}\\
 = \eta_\nu - \chi_\nu\inu.\nonumber
\ee
In writing Eq.~\ref{rteob}, we have retained the first term, which
accounts for the explicit time dependence of the radiation field in the
co-moving frame. We have also retained the acceleration term,
$\frac{\partial\beta}{\partial t}$.  Both terms are
of $O(\beta)$ when compared to other terms in the equation, such as
the $\beta/r$ terms, and hence, are of
$O(\beta^2)$ on  a
fluid flow timescale and can be dropped
\cite[]{castor72,buchler79,mih80,found84}. Upon doing so, one derives
the time-independent (or quasi-static) transfer equation in the
co-moving frame:
\be
(\mu + \beta) \frac{\partial\inu}{\partial r} + (1-\mu^2)\left[ \mu
(\frac{\beta}{r} 
- \frac{\partial\beta}{\partial r}) 
+ \frac{1}{r}
\right]\frac{\partial\inu}{\partial \mu}\label{rteobqs}\\
+ \left[
\mu^2(\frac{\beta}{r} - \frac{\partial\beta}{\partial r}) -
\frac{\beta}{r}\right]\left\{ \frac{\partial\inu}{\partial \ln\nu} -
3\inu\right\}\nonumber\\
 = \eta_\nu - \chi_\nu\inu.\nonumber
\ee

To further simplify the equation and to help elucidate the fundamental
physics, let us restrict ourselves to consideration of homologous flows:
$\beta \propto r$. In this case Eq.~\ref{rteobqs} becomes:
\be
&&(\mu + \beta) \frac{\partial\inu}{\partial r} +
\frac{(1-\mu^2)}{r}\frac{\partial\inu}{\partial \mu}
- 
\frac{\beta}{r}\left\{\frac{\partial\inu}{\partial \ln\nu} -
3\inu\right\}\nonumber\\
&\quad& = \eta_\nu - \chi_\nu\inu.\label{homorteob}
\ee
In order to identify the physical significance of the terms, it is
useful to compare this equation to its static counterpart:
\be
\mu \frac{\partial\inu}{\partial r} +
\frac{(1-\mu^2)}{r}\frac{\partial\inu}{\partial \mu}
= \eta_\nu - \chi_\nu\inu.\label{staticrte}
\ee
Comparing Eqs.~\ref{homorteob} and~\ref{staticrte}, the physical meaning
of the terms is apparent: $\beta \frac{\partial\inu}{\partial r}$ is
the advection term, $
{\beta}/{r}\frac{\partial\inu}{\partial
\ln\nu}$ represents the Doppler shift, and
$-3\beta/r\inu$ describes the effect of aberration.

It is clear that all three terms are of O($\beta$) and must be
retained to have a consistent treatment in the co-moving frame. 
The $O(\beta)$ transport equation is more difficult to solve because
the characteristics, which are simply parallel lines of constant impact
parameter when the advection term, $\beta
\frac{\partial\inu}{\partial r}$, is neglected, become curved lines; 
therefore, the reflection symmetry is lost \cite[]{mkh76b,mih80}. This
is because 
 the material is moving, ``sweeping up'' 
radiation, causing   the characteristics to be curved. Therefore, one
can no longer use reflection symmetry to integrate the solution only along
outgoing rays. One must integrate along both incoming and outgoing
rays \cite[]{mkh76b,mih80,phhs392}. \cite{mkh76b} examined the magnitude of
the advection and aberration terms and estimated that they are of
order $5\beta$ and that the advection term is more
important than the aberration term.

Let us now examine the moments of Eq.~\ref{homorteob}. The zeroth
moment is: 
\be
\beta\frac{\partial\jnu}{\partial r} + \frac{1}{r^2}\frac{\partial
(r^2\hnu)}{\partial r}
+\frac{\beta}{r}(3\jnu - \frac{\partial\jnu}{\partial\ln\nu})
 = \eta_\nu - \chi_\nu \jnu, \label{zero}
\ee
and the first moment is:
\be
\frac{\partial\knu}{\partial r} &+& \frac{\beta}{r^2}\frac{\partial
(r^2\hnu)}{\partial r} - \frac{(\jnu-3\knu)}{r}
+\frac{\beta}{r}(\hnu - \frac{\partial\hnu}{\partial\ln\nu})\nonumber\\
 &=& - \chi_\nu \hnu. \label{one}
\ee
The Eddington moments are given by:
\be
J_\nu &=& \frac{1}{2} \int_{-1}^{1} \inu\,d\mu\nonumber\\
H_\nu &=& \frac{1}{2} \int_{-1}^{1} \mu\inu\,d\mu\\
K_\nu &=& \frac{1}{2} \int_{-1}^{1} \mu^2\inu\,d\mu\nonumber
\ee
where $\mu=\cos\theta$.
When advection is neglected, the moment equations become:
\be
&\displaystyle \frac{1}{r^2}\frac{\partial
(r^2\hnu)}{\partial r}&
+\frac{\beta}{r}(3\jnu - \frac{\partial\jnu}{\partial\ln\nu})
= \eta_\nu - \chi_\nu \jnu, \label{zeronoad}\\
&\displaystyle\frac{\partial\knu}{\partial r}&  - \frac{(\jnu-3\knu)}{r}
+\frac{\beta}{r}(3\hnu - \frac{\partial\hnu}{\partial\ln\nu})\nonumber\\
 &=& - \chi_\nu \hnu, \label{onenoad}
\ee
i.e., the gradient of the energy density is absent from
the zeroth moment equation, and the divergence of the flux is no
longer included 
in the first moment equation.
Integrating the moment equations (Eqs.~\ref{zero} and \ref{one}) over
frequency, and assuming that radiative equilibrium holds, i.e., that energy is
conserved or total emission equals  total absorption [$\int_0^\infty (\eta_\nu
- \chi_\nu \jnu)\,d\nu = 0$], one obtains for the zeroth moment: \be
\frac{D\,J}{D\,t} + \frac{1}{r^2}\frac{\partial (r^2\,H)}{\partial r}
+\frac{4\beta}{r}J = 0,\label{zerofi} \ee where we have restored the
time derivative:
\be
\frac{D}{Dt} = \frac{\partial}{\partial t} + \beta\frac{\partial}{\partial r}.
\ee

It has been suggested \cite[]{eastpin93} that one can correct for 
neglecting the advection term and include the effects of the
radiation field time
dependence on a radiation flow timescale by
using Eq.~\ref{zerofi} and by arbitrarily setting the co-moving
luminosity ($ = r^2\,H $) to be constant. In this case, Eq.~\ref{zerofi}
becomes: \be \frac{D\,J}{D\,t} = -\frac{4\beta}{r}J. \label{pefi} \ee
This is interpreted as an operator equality:
\be
\frac{D}{Dt} = -\frac{4\beta}{r}.\label{dumb}
\ee
When Eq.~\ref{dumb} is substituted into the radiation
transport equation, the transport equation  becomes: 
\be \mu
\frac{\partial\inu}{\partial r} +
\frac{(1-\mu^2)}{r}\frac{\partial\inu}{\partial \mu} -
\frac{\beta}{r}\left\{\frac{\partial\inu}{\partial \ln\nu} +
\inu\right\} = \eta_\nu - \chi_\nu\inu.\label{eprte} \ee Comparing
Eq.~\ref{eprte} to Eq.~\ref{homorteob}, we see that
this scheme is equivalent to making the quasi-static approximation,
neglecting advection, and changing the sign and coefficient of the
aberration term, which is unphysical.

\begin{figure*}
\begin{center}
\leavevmode
\psfig{file=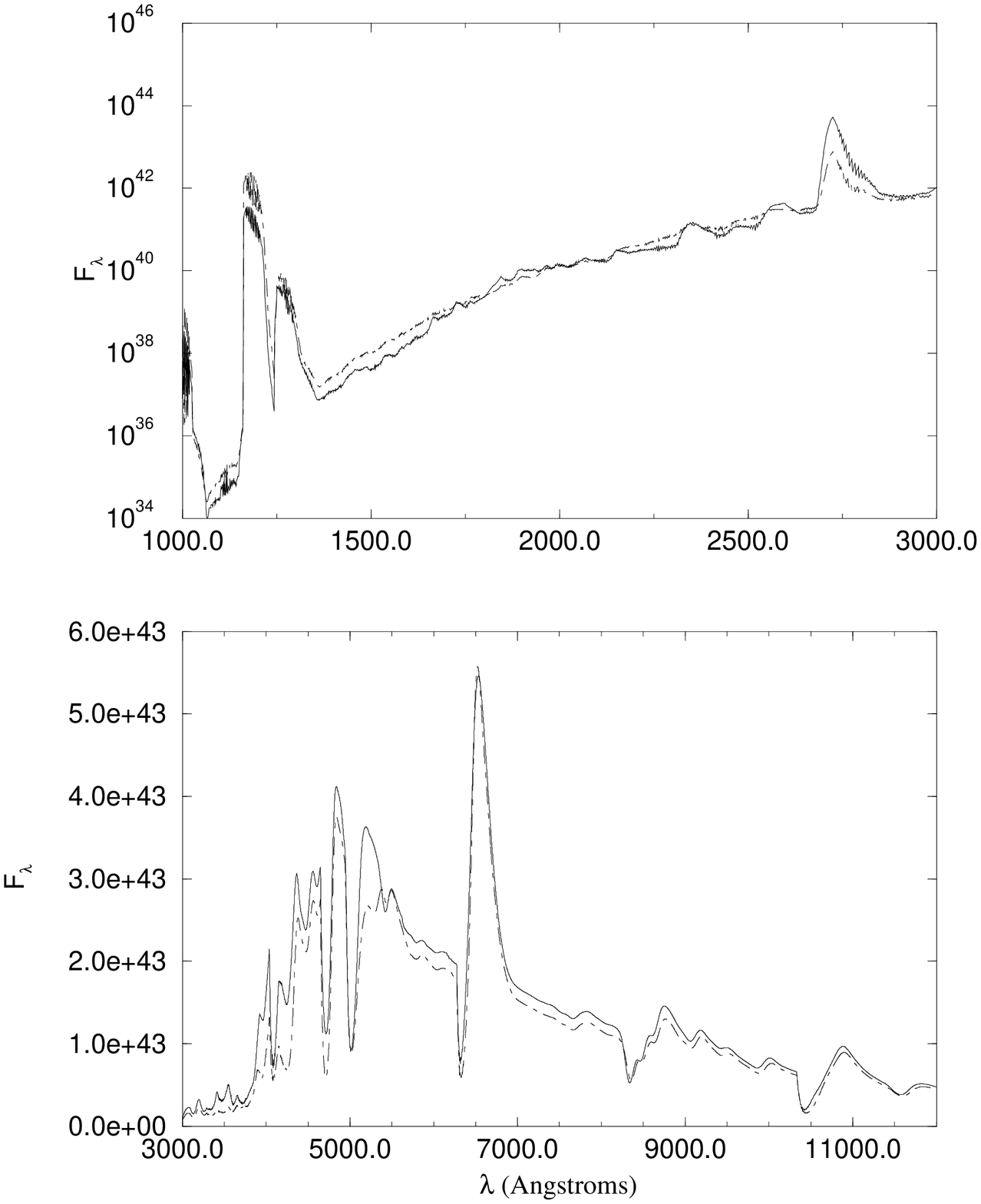,height=0.70\textheight,angle=0}
\end{center}
\caption{\label{sn87a_d13_sp}The spectrum produced by a full
calculation, which fit SN 1987A at 13 days past explosion (solid
line), is compared to one with the same parameters in which advection
is neglected (dot-dashed line). Both calculations are in radiative
equilibrium.}
\end{figure*}

\begin{figure*}
\begin{center}
\leavevmode
\psfig{file=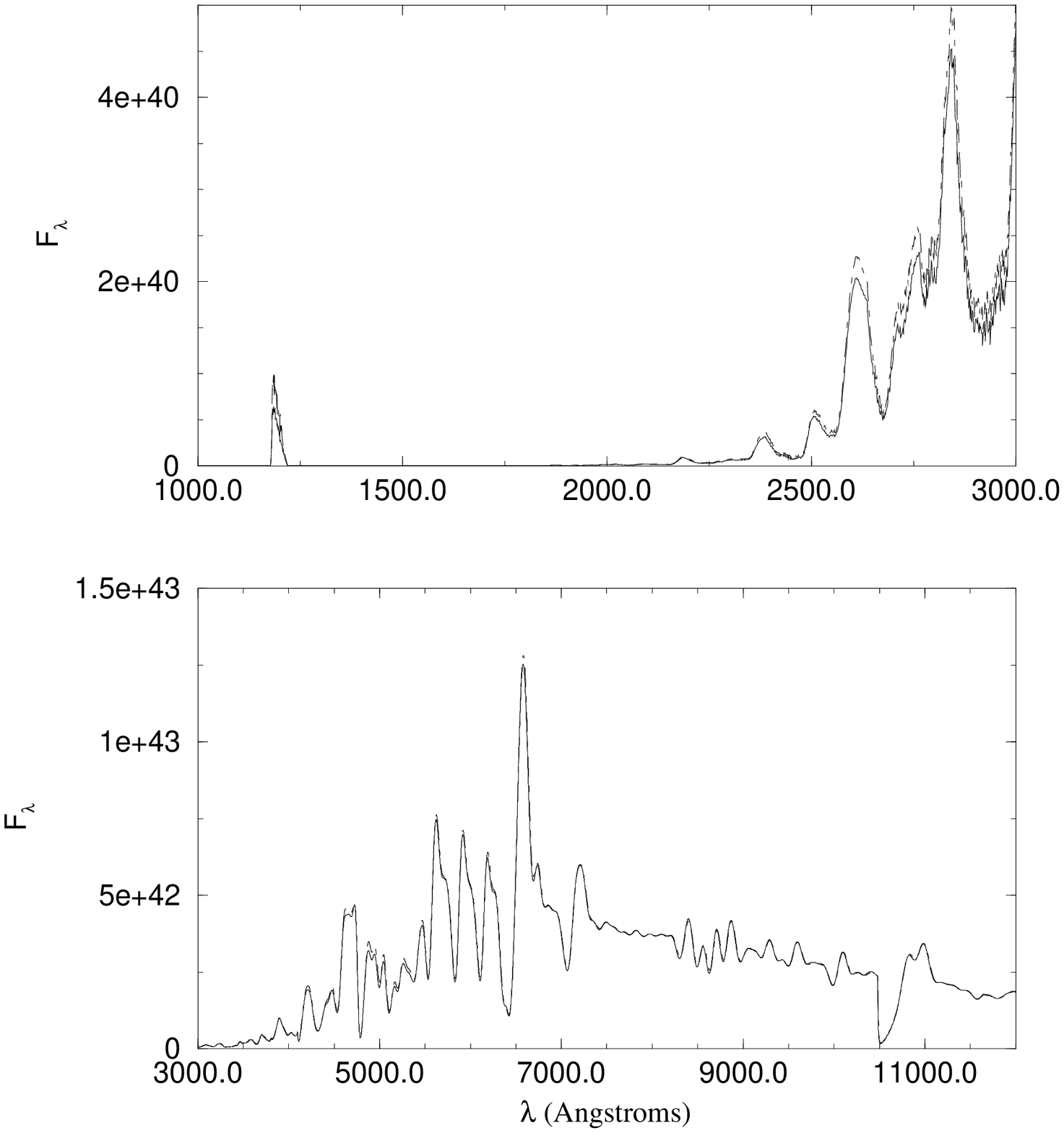,height=0.70\textheight,angle=0}
\end{center}
\caption{\label{sn87a_d31_sp}The spectrum produced by a full
calculation, which fit SN 1987A at 31 days past explosion (solid
line), is compared to one with the same parameters in which advection
is neglected (dot-dashed line). Both calculations are in radiative
equilibrium.}
\end{figure*}

\begin{figure*}
\begin{center}
\leavevmode
\psfig{file=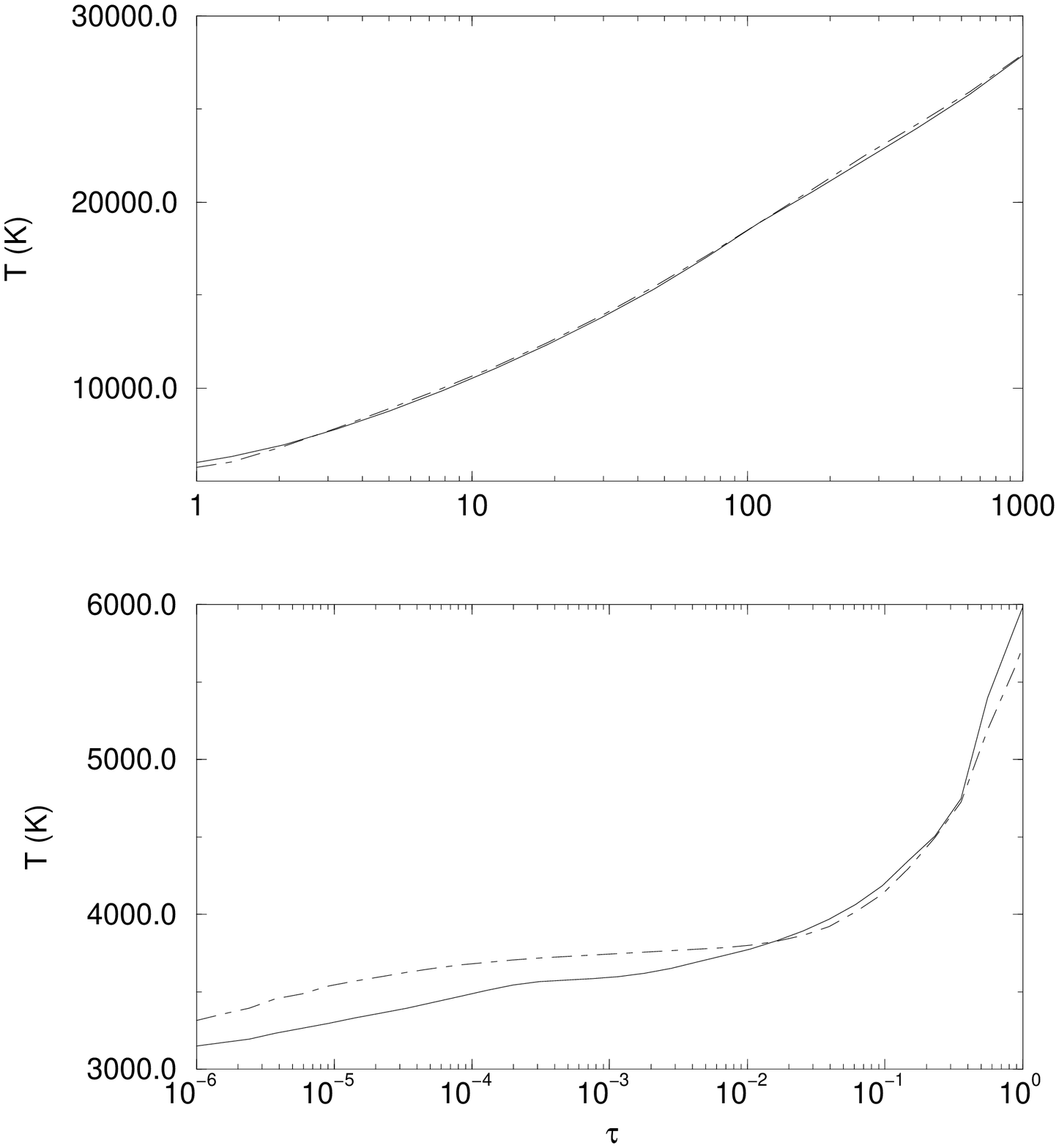,height=0.65\textheight,angle=0}
\end{center}
\caption{\label{day31_temps}The temperature profiles for the models
plotted in Fig.~\protect\ref{sn87a_d13_sp} are compared.}
\end{figure*}

\begin{figure*}
\begin{center}
\leavevmode
\psfig{file=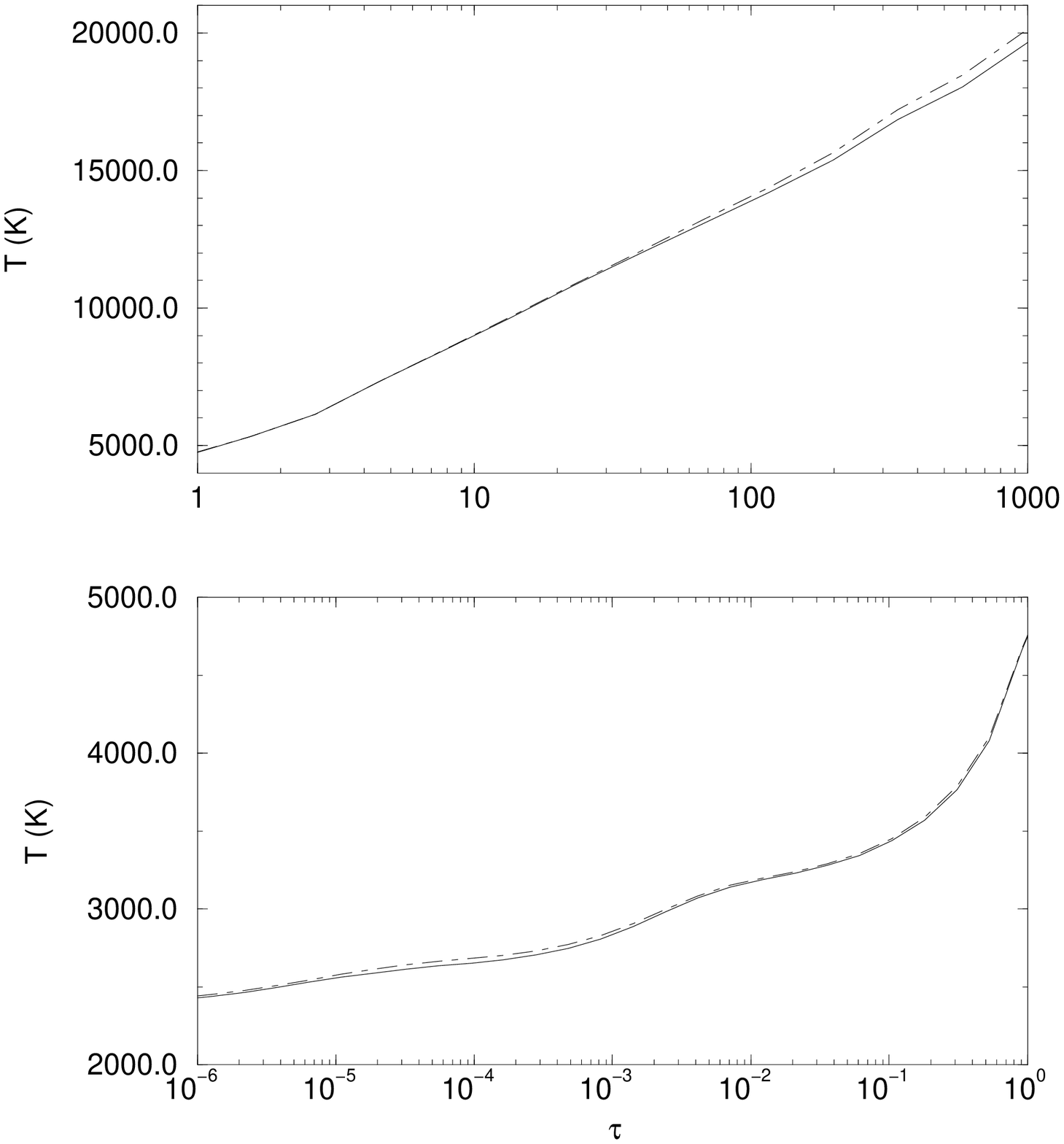,height=0.65\textheight,angle=0}
\end{center}
\caption{\label{day13_temps}The temperature profiles for the models
plotted in Fig.~\protect\ref{sn87a_d31_sp} are compared.}
\end{figure*}

\begin{figure*}
\begin{center}
\leavevmode
\psfig{file=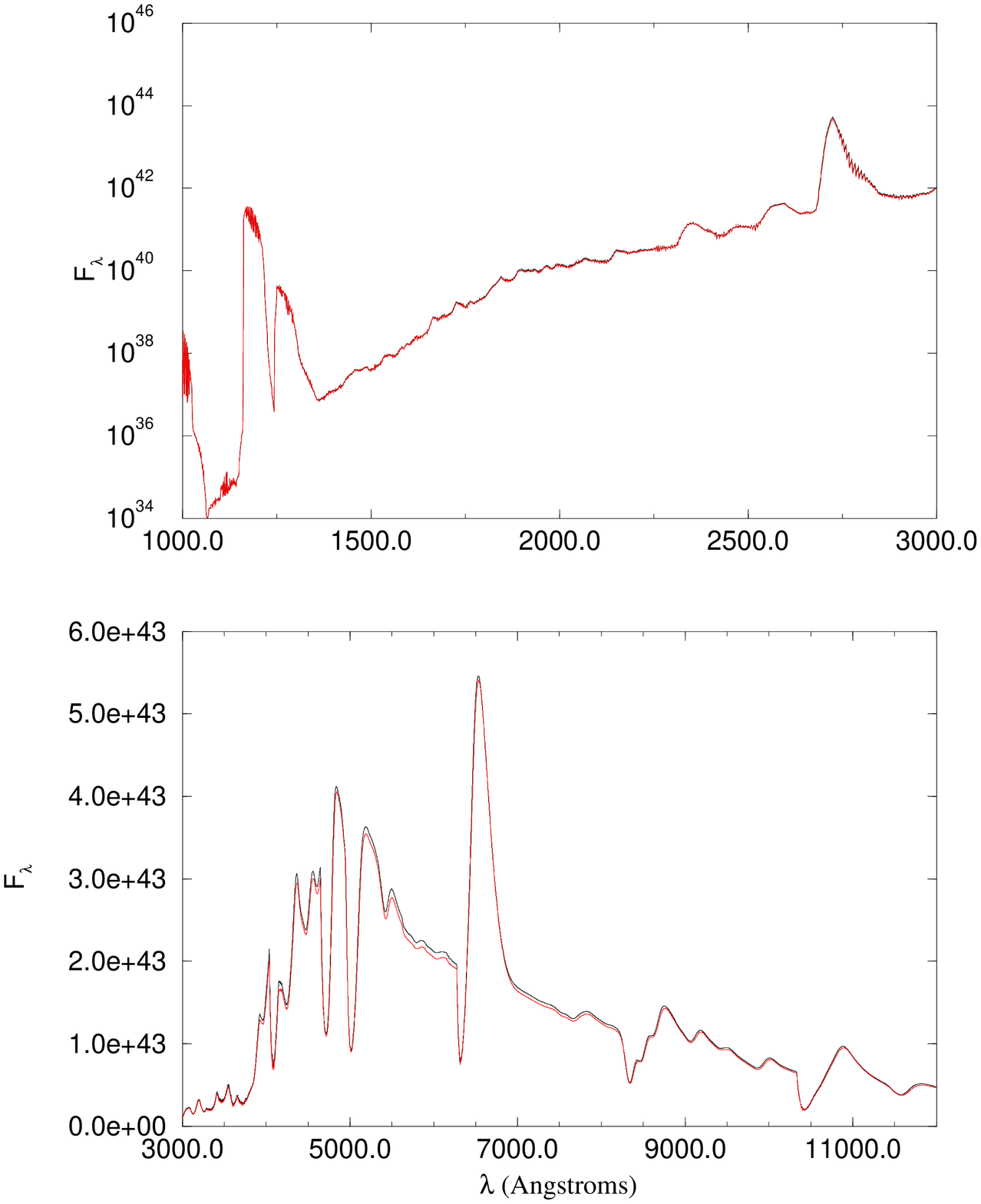,height=0.70\textheight,angle=0}
\end{center}
\caption{\label{fixstruct_d13_sp}The spectrum produced by a full
calculation, which fit SN 1987A at 13 days past explosion (solid
line), is compared to one with the same parameters in which advection
is neglected (dotted line). The structure is fixed to be that of the
full calculation.}
\end{figure*}

We have compared simulations with and without the advection term
for two models that provide reasonable fits to SN~1987A
at 13 and 31 days after explosion. These calculations were performed
using version 5.5.9 of the general radiative transfer code, {\tt
PHOENIX}, developed by Hauschildt
\cite[]{phhs392,phhre92,phhcas93}. This code accurately solves the
fully relativistic transfer equation, Eq.~\ref{fullrte}, in the 
quasi-static approximation, $\frac{\partial \inu}{\partial t} = 0$
(Note contrary to what is incorrectly stated
by H{\"o}flich \etal, in  this
volume, we do 
not  solve a simple non-relativistic transport equation with the Rybicki
method, but we indeed solve the full special relativistic,
spherically symmetric radiative 
transfer equation for lines and continua with an operator splitting
scheme based on a short characteristic method with non-local,
adjustable approximate $\Lambda$-operator).
 The
model parameters are given in Table~\ref{params} \cite[for a
discussion of the model parameters,
cf.][]{b93j3}. For the day-13 spectrum, Figure~\ref{sn87a_d13_sp}
compares the spectra of 
a  calculation that includes advection with one  that does not,
 while Figure~\ref{sn87a_d31_sp} displays the same for 
day-31.  The differences are about the size predicted by
\cite{mkh76b}, with the effects being more apparent in the faster 
day-13 model than in the much slower day-31 model. 
Figures~\ref{day31_temps} and~\ref{day13_temps} show comparisons of
the temperature profiles for both models.   Neglecting advection
alters the temperature structure, which can be interpreted as
resulting from 
the change in the relations between the moments (compare
Eqs.~\ref{zero} and \ref{one} with Eqs.~\ref{zeronoad} and
\ref{onenoad}).  Figure~\ref{fixstruct_d13_sp} illustrates that this
is the most important effect of neglecting 
advection.
The temperature structure and departure coefficients are kept fixed
in order to 
 only alter the transport equation. In this case, the emergent
spectra are much more similar than those in Fig.~\ref{sn87a_d13_sp}.
From these results, it is 
clear that the effects of advection should not be neglected in models
of supernovae where the characteristic velocities are larger than
$\approx 5000$~\kmps. For systems where velocities are lower than
$\sim 2000$~\kmps,\ 
advection may be neglected with reasonable accuracy. 

\begin{table}
\begin{center}
\begin{tabular}{lll}
 & Day 13 & Day 31\\
\hline\\
\Teff (K) & 5400& 4400\\
\vno (\kmps)& 5500&1700\\
\Rzero (cm)&$6.1 \times 10^{14}$&$4.5 \times 10^{14}$\\
N& 6 & 6
\end{tabular}
\end{center}
\caption{\label{params}$\Teff$,
\Rzero, $\vno$ are the ``effective'' temperature, radius, and velocity
at the reference point $\tau_{\rm std}=1$;
N is the power-law
density index.}
\end{table}

\section{Quasi-Static Approximation}

We can estimate the effects of making the  quasi-static
approximation by examining the radiation transfer equation.  For
clarity, we again restrict ourselves to $O(\beta)$ and homologous
flows. Restoring the time derivative in Eq.~\ref{homorteob}, we have:

\vbox{
\be
\frac{\partial\inu}{\partial t}& + &(\mu + \beta)
\frac{\partial\inu}{\partial r} + 
\frac{(1-\mu^2)}{r}\frac{\partial\inu}{\partial \mu}\nonumber\\
&-& 
\frac{\beta}{r}\left\{\frac{\partial\inu}{\partial \ln\nu} -
3\inu\right\}\nonumber\\
&\quad& = \eta_\nu - \chi_\nu\inu.\label{homorteobwt}
\ee
}

In order to solve this equation numerically, we would replace the time
derivative with the difference:
\be
\frac{\partial\inu}{\partial t} = \frac{\inu - \inu^n}{\delta
t},\label{tderiv}
\ee
where $\inu^n$ is the intensity evaluated at the previous time $t^n$,
$\inu$ is the 
intensity at the current time $t^{n+1}$, and $\delta t = t^{n+1}-t^n$.
Inserting this expression into Eq.~\ref{homorteobwt}, and moving the
time derivative to the right-hand side of the transfer equation, we obtain:
\be
(\mu + \beta) \frac{\partial\inu}{\partial r} &+&
\frac{(1-\mu^2)}{r}\frac{\partial\inu}{\partial \mu}
- 
\frac{\beta}{r}\left\{\frac{\partial\inu}{\partial \ln\nu} -
3\inu\right\}\nonumber\\
 &=& (\eta_\nu + \frac{\inu^n}{\delta t}) - (\chi_\nu +
\frac{1}{\delta t})\inu,\label{rtetd} 
\ee
which shows that the time derivative term can be viewed as an additional
source and sink of radiation. We can estimate the size of the error
made in the quasi-static approximation
by
examining the ratio $\chi^{-1}/(c \delta t)$, where we have restored
the explicit $c$. In supernovae, the
natural timescale is the age of the object, $t = R/v$. We may
estimate that $\chi
\approx \tau/R$, where $\tau$ is an appropriate optical depth. Then,
the ratio becomes:
\be
\frac{\chi^{-1}}{c \delta t} \approx \frac{\beta}{\tau}.\label{err}
\ee
For Type Ia supernovae at maximum light, the continuum extinction
optical depth is 
about 10, and $\beta \sim 1/30$. So the error is at most 0.3\%;
small compared to errors in the atomic physics. This error will
be considerably smaller for other types of supernovae, which are
optically thick for longer times. In fact, Eq.~\ref{err} is
an overestimate of the error because we have neglected the source term 
$\inu^n/(c\delta t)$, which counteracts the extra sink term.

Claims (Eastman, this volume) that in order to account for ``old
photons'', time dependence must be included in the transfer equation
are not correct.  The effects of both dynamic and static diffusion are
included in radiation hydrodynamics {\em without\/} including the
time-dependent term in the transfer equation \cite[]{found84}. The
effects
of departures from radiative equilibrium can be included in our modeling.

\section{NLTE Effects}

\begin{figure*}
\begin{center}
\leavevmode
\psfig{file=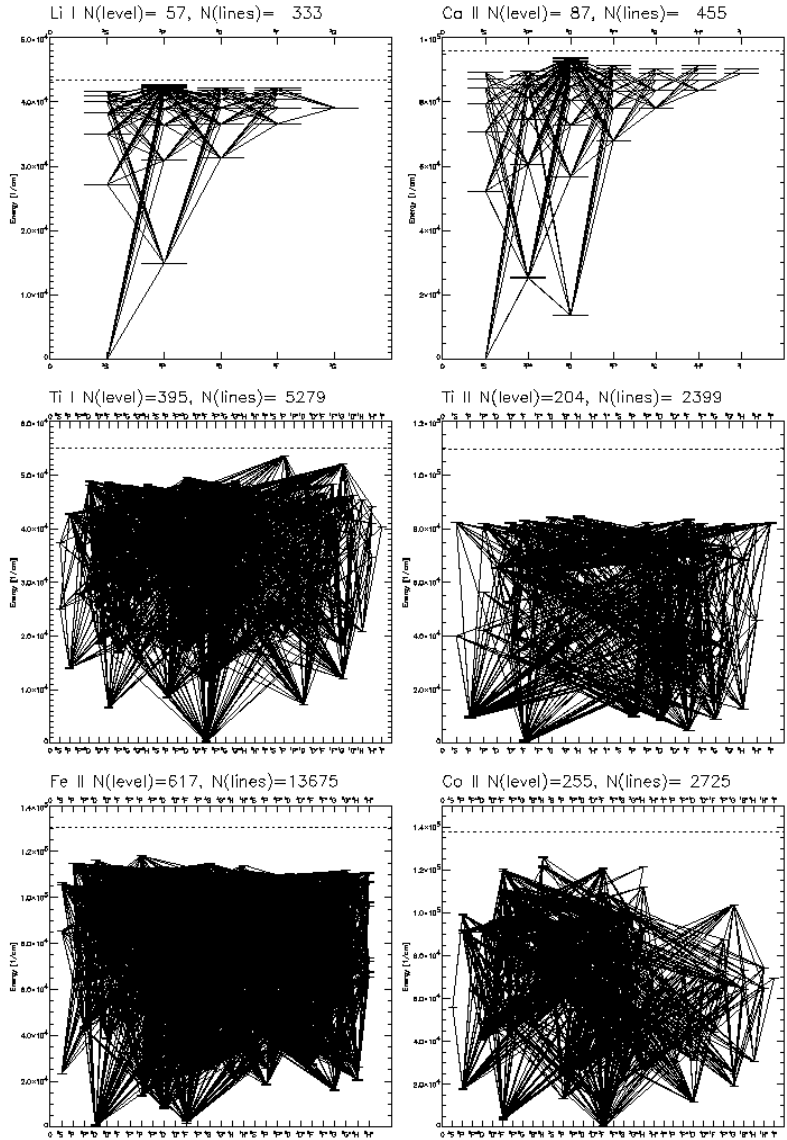,height=0.85\textheight,angle=0}
\end{center}
\caption{\label{grot} Grotrian diagrams for Li~I, Ca~II, Ti~I, Ti~II,
Fe~II, and Co~II model atoms used in the calculations. Only the
``primary'' radiative transitions are displayed; see
\protect\cite{hbfe295} and \protect\cite{phhnovfe296} for more details.
}
\end{figure*}

Figure~\ref{grot} displays the model atoms for Li~I, Ca~II, Ti~I,
Ti~II, Fe~II, and Co~II used in {\tt PHOENIX}. In addition H~I, He~I,
He~II, Mg~II, Ne~I, and O~I are also treated in NLTE. In the very near
future we will add C~I-IV, N~I-VI, O~II--VI, Si~II--III, and
S~II--III.  In particular, our large Fe~II model atom \cite[617 levels,
13675 primary transitions; see][for more details]{hbfe295} allows us to
determine the importance of the size of model atoms and of the
validity of common
assumptions that are often made in handling the millions of secondary
transitions that must be included in order to correctly reproduce the
UV line blanketing.  We find that large changes in luminosity (see
Pinto, this volume) can be avoided by treating the secondary lines
with a small but non-zero thermalization parameter \cite[]{snefe296},
as required by physical considerations.

\section{Sobolev Approximation}

\begin{figure}
\psfig{file=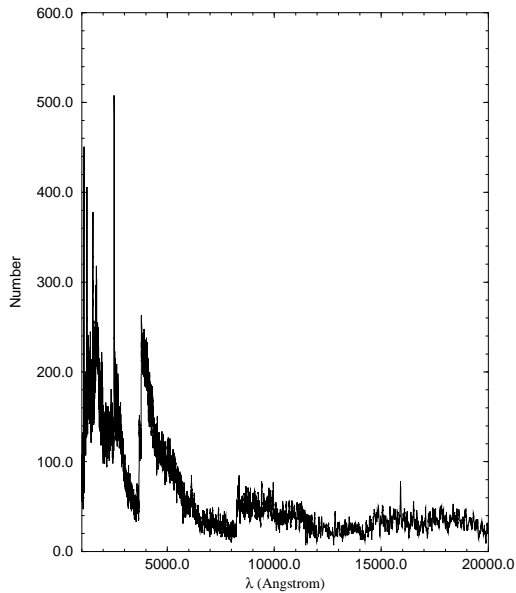,height=0.5\textheight,angle=0}
\caption{\label{sob} The number of overlapping lines in a range of 6
Doppler widths around any 
given wavelength, as a function of wavelength, for the day-13 SN
1987A model. Only lines stronger than $10^{-3}$ times the local b-f
continuum are included. Clearly, the Sobolev assumption  that
individual lines do not overlap is {\em not\/} fulfilled.}
\end{figure}

The Sobolev approximation developed by \cite{sob60} and
\cite{castor70} and extended by
\cite{humryb85,humryb92} and \cite{jeff89,jeff90,jeff95,jeff96} has
proven extremely 
valuable in providing line identifications and minimum and maximum
velocities in supernovae
\cite[]{bran81b,branch81b85,jb90,jbfn91,jeffetal92,hst93j,kir92a,fil91bg92},
because it allows one to calculate line profiles without solving the
transfer equation, it is nearly analytic, and quite convenient. The
above analyses were concerned with identifying strong lines, and
continuum effects were neglected. More recently the Sobolev
approximation has been used to solve the rate equations for detailed
model atoms including continua \cite[][H{\"o}flich \etal, this
volume]{eastpin93,hofsn94d}. However, 
because the escape probability is derived
by neglecting the effects of neighboring lines, it is only valid
for isolated lines, and is invalid when there are many weak
overlapping lines \cite[]{castor70,ryb84,avrloes87}. This is likely to
be the case in the UV, where line blanketing is severe. \cite{ryb84}
also has discussed that escape probability methods such as the Sobolev
approximation are inaccurate at small line optical depths,
particularly when there are many overlapping lines. Since the source
function predicted by the Sobolev approximation at the surface is
incorrect by a factor of $\sqrt{\epsilon}$, where $\epsilon$ is the
line thermalization parameter, the value of $J$ found by the formal
solution will also be in error, which in turn will lead to errors in
the rate equations. In Figure~\ref{sob} we display the number of
overlapping lines in a range of 6 intrinsic Doppler widths  around
any given wavelength, as a function of wavelength, for the
day-13 SN 1987A model, which has a statistical or micro-turbulent
velocity of $\xi = 50$~\kmps. The lines are said to overlap if, for any
particular line, another line has its line center $\pm 6$ intrinsic
line-widths from the reference line.  The Doppler widths are calculated
at deepest depth point in the model. As expected, in the UV the mean
number of overlapping lines is typically around 100, and can be as
large as 500.  
 This implies that the radiative transfer in SN (and nova) atmospheres 
must explicitly include the effects of  overlapping lines and 
continua. Otherwise, the radiative rates for these transitions would be 
incorrect, particularly in the outer parts of the atmosphere where 
ionization corrections are most important. Although this requires a very 
fine wavelength grid for the model calculations, detailed models can be 
computed using modern numerical techniques on even moderately sized 
workstations. 
Thus,  the Sobolev approximation cannot be
used in detailed NLTE calculations for SNe, because the radiative rates
calculated in this approximation are inaccurate. Similar results have been
obtained by \cite{phhnov95} in nova model atmosphere calculations.

For pure hydrogen atmospheres, \cite{duschetal95} found good
agreement between the non-relativistic 
Sobolev approximation and non-relativistic co-moving frame full transport
calculations,  which shows that the
Sobolev approximation is accurate for well separated lines such as
the Balmer lines.  However, this situation is not reproduced
in most spectral regions, and therefore, the Sobolev approximation is of
limited use for detailed  modeling of SN or nova envelopes.

\section{Expansion Opacities}

Radiation hydrodynamic calculations of supernova light curves require
accurate fluxes, and it has long been realized that the static Rosseland
mean opacity does not produce an accurate flux in moving atmospheres
\cite[]{karpetal77}. The work of \cite{karpetal77}
provided an approximate formula for the Rosseland
mean opacity in the observer's frame. However, nearly all
radiation hydrodynamics  
calculations are performed in the co-moving frame; hence, a co-moving
formulation is required.

We have derived the Rosseland mean opacity in the co-moving frame to
$O(\beta)$ \cite[]{bhm95}. Let us first recall that the static Rosseland mean,
$\chi_R^0$, is
given by:
\be
\frac{1}{\chi_R^0} = (4\frac{\sigma}{\pi}T^3)^{-1} \int_0^\infty \chi_\nu^{-1}
\frac{dB_\nu}{dT}\,d\nu.\label{rossstat}
\ee
To derive the non-static Rosseland mean, we will assume homologous flows.
In addition, we make the Eddington approximation,
$\knu=1/3\jnu$, implying that the co-moving radiation field is close to
isotropic, which is an excellent approximation in the diffusive regime (large
optical depth) because the radiation is collision dominated
\cite[]{pomnotes82}. 

We find that the 
co-moving  Rosseland
mean opacity to $O(\beta)$ is given by:
\be
\frac{1}{\chi_R^\beta}
 \equiv (\chi + \frac{2\beta}{r})^{-1}\label{rossbeta}
\ee
in the gray case, and:
\be
\frac{1}{\chi_R^\beta} =\mbox{\hspace{2.3in}}&\quad&\nonumber\\
(4\frac{\sigma}{\pi}T^3)^{-1} \int_0^\infty \chi_\nu^{-1}
[1-\frac{\beta}{r\chi^0_R} (1-\frac{\partial}{\partial\ln\nu})]
\frac{dB_\nu}{dT} \,d\nu,&\quad&\\
= \frac{1}{\chi^0_R}\left[{1 - \frac{\beta}{\chi^0_Rr} +
\frac{\beta}{r}(4\frac{\sigma}{\pi}T^3)^{-1} \int_0^\infty 
\chi_\nu^{-1} \frac{\partial^2 B_\nu}{\partial T\partial
\ln\nu}\,d{\nu}}\right],&\quad&\label{rossbet}
\ee
in the non-gray case. In deriving Eq.~\ref{rossbet}, we have used
$\frac{\partial B}{\partial r} = 4 (\sigma/\pi) T^3\frac{\partial
T}{\partial r}$. 

It follows that the co-moving multi-group flux to be used in
radiation hydrodynamics is given by a Fick's law
diffusion equation:
\be
H_\nu^\beta = -\frac{1}{3\chi_R^\beta} \frac{\partial B_\nu}{\partial r}.
\ee
We emphasize that $\chi_\nu$ in Eq.~\ref{rossstat}
 and \ref{rossbet}
contains contributions from continua, lines, and scattering opacities, and
nowhere have we had to treat lines differently from continua.

\begin{figure*}
\begin{center}
\leavevmode
\psfig{file=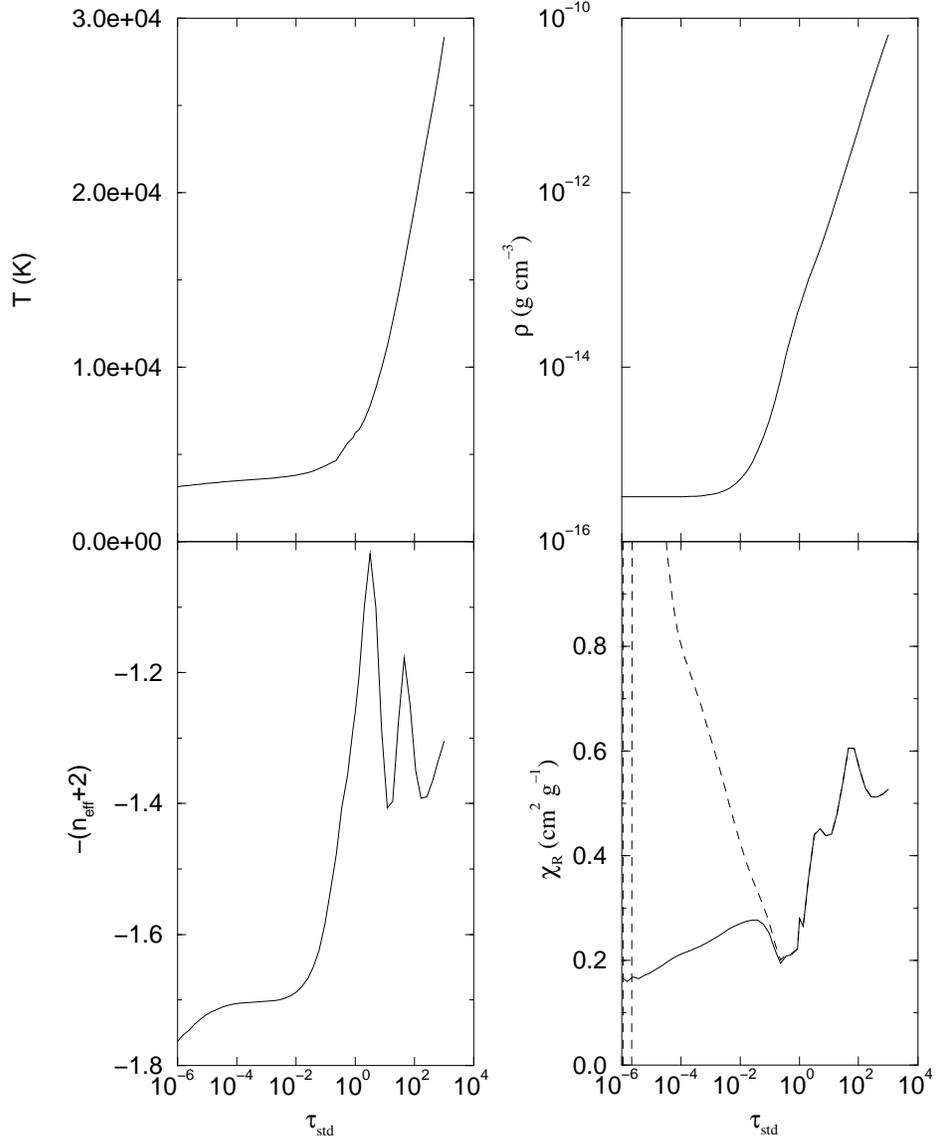,height=0.85\textheight,angle=0}
\end{center}
\caption{\label{opacs} The 
 effective
value of n, the density, the temperature, $\chi^0_R$ (solid line), and
$\chi^\beta_R$ (dashed line),
as functions of $\tau_{\rm std}$, for the day-13 model. At very low
optical depths, 
the formula breaks down.
}
\end{figure*}

In the case that the opacity may be approximated by a power-law, $\chi_\nu
\propto \nu^{-n}$, the last integral in Eq.~\ref{rossbet}
may be evaluated by an integration by parts, yielding:
\be
\frac{\beta}{r}(4\frac{\sigma}{\pi}T^3)^{-1} \int_0^\infty
\chi_\nu^{-1} \frac{\partial^2 B_\nu}{\partial T\partial
\ln\nu}\,d{\nu} = -(n+1) \frac{\beta}{r\chi^0_R},
\ee
and therefore:
\be
 \chi_R^\beta \approx
{\chi^0_R}(1-(n+2)\frac{\beta}{r\chi^0_R})^{-1}.\label{neffdef} 
\ee
We have calculated the correction factor for atmospheres appropriate
to Type II supernovae. As an illustrative case, Figure~\ref{opacs}
displays  the density, temperature, $\chi^0_R$,  $\chi^\beta_R$,
and the effective 
value of n (that is, the value of n one obtains from Eq.~\ref{neffdef}
using the exact values of $\chi^0_R$ and  $\chi^\beta_R$) as functions
of $\tau$ for the day-13 model. 
As expected from Eq.~\ref{rossbeta}, the largest correction occurs at
low optical depth (the formula breaks down at very small optical
depths since the assumptions used to derive it, i.e., LTE and the
Eddington approximation, are not fulfilled), and the correction is
essentially irrelevant at high optical 
depths, where $1/\chi_R^0 << \beta/r$. 

\section{Conclusions}

We have shown that advection cannot be neglected in the co-moving
solution of the radiation transport equation. Its main influence is on
the temperature structure, through the term it adds to the equation of
radiative equilibrium. The errors made in neglecting advection scale
with the velocity; while it may be acceptable to neglect advection
in systems where the velocities are $< 2000$~\kmps, such as in hot stars,
novae with low wind velocities (e.g, Nova Cas 1993), and Type II
supernovae at late times, it {\em cannot} be neglected for supernovae
at early times and novae with high wind velocities (e.g., Nova Cygni
1992).

We have also shown that the Sobolev approximation is likely to be
invalid for weak lines in the co-moving frame, since many of these
lines overlap.

We have derived an approximate expression [good to $O(\beta)$] for the
Rosseland mean opacity that can be used in radiation hydrodynamics
calculations. The  Doppler shift is fully
accounted for in this approximation. Our formula shows that, at large
optical depths, the static Rosseland mean is accurate, and hence, for
all radiation hydrodynamics calculations that use flux-limited diffusion, the
static approximation is excellent.

\section*{Acknowledgments}
We thank Sergej Blinnikov, David Branch,  and Sumner
Starrfield for 
helpful discussions. This work
was supported in part by NASA grant NAGW-2999, and a NASA LTSA grant
to ASU, and by NSF grant AST-9417242.  A.M. is supported in part at
the University of Tennessee 
under DOE contract DE--FG05--93ER40770, and at the Oak Ridge National
Laboratory, which is managed by Lockheed Martin Energy Systems
Inc. under DOE contract DE--AC05--84OR21400. Some of the calculations
in this paper were performed at the NERSC, supported by the U.S. DOE,
and at the San Diego Supercomputer Center, supported by the NSF; we
thank them for a generous allocation of computer time.


\end{document}